# A Multidisciplinary Undergraduate Nanoscience and Nanotechnology Program at the University of North Dakota

Naima Kaabouch, Deborah Worley, Matt Cavalli, Kanishka Marasinghe,

Nuri Oncel, David Pierce, Brian Tande, Julia Zhao

**Abstract**

This paper describes some of the results of a National Science Foundation Nanotechnology Undergraduate Education project that aims to establish a nanoscience and nanotechnology program at the University of North Dakota. The goal is to generate new interest in nanoscience and nanotechnology among STEM students and prepare them with the knowledge and skills necessary for the next generation of graduates to compete in the global market and contribute to the nanoscience and nanotechnology field. The project explored several aspects of student learning, including students' motivations for investigating nanotechnology through interdisciplinary coursework. To collect this information, a survey was administered to students who enrolled to two nanoscience and nanotechnology courses. Data collected from the survey will be used to improve the design and delivery of future courses as part of constructing a complete nanoscience and nanotechnology curriculum.

## I    Introduction

A multidisciplinary team has developed a nanoscience & nanotechnology program at the University of North Dakota (UND) with the support of the National Science Foundation (NSF) Nanotechnology Undergraduate Education (NUE) program [1-6]. This NSF sponsored project which started in September 2014 aimed at fostering new interest in nanoscience and nanotechnology among UND STEM students from a multidisciplinary perspective. This program has unique features as it added experimental learning to class lectures. The outcomes of the project were 1) development of two nanoscience and nanotechnology (NSNT) courses, 2) development of lab activities integrated in the first course, and 3) development of activities for American Indian students.

The two courses were offered by faculty from Electrical Engineering, Mechanical Engineering, Chemical Engineering, Chemistry, and Physics. This multidisciplinary delivery enables STEM students to understand the multidisciplinary nature of the NSNT field. This approach encourages collaborative and multidisciplinary learning for the students and helps them acquire the knowledge and skills necessary to compete in the global market and to contribute to the NSNT field in an environment that is reflective of today's workplace.

The project explored several aspects related to student learning, including students' motivations for investigating nanotechnology through interdisciplinary coursework. To collect this information, a NSNT Survey was administered to students who enrolled to the two NSNT courses. Data collected from the survey will be used to improve the design and delivery of future nanoscience and nanotechnology (NSNT) courses as part of constructing a complete nanoscience and nanotechnology curriculum.

This paper describes the 2 courses and the results of the survey. Other elements of the project have been covered in previous papers [1-3] and others will be covered in subsequent papers. This paper is organized as follows: Section 2 gives an overview on the two NSNT courses, the teaching methods, and the lab activities; and Sections 3 and 4 gives and discuss the results of the survey administered to students in 2016 and 2017. Finally, Section 5 offers our conclusions from the research effort.

## II   Courses

For this project, the collaborators developed and offered two NSNT courses that satisfy elective requirements for mechanical, chemical, and electrical engineering Bachelor of Science (BS) degrees as well as for chemistry and physics BS degrees.

The first course, "Nanoengineering and Nanoscience" (cross-listed course numbers: CHEM 431; EE490; ME490; PHYS492), was offered during the fall semesters of 2015 and 2016. It covered the fundamentals of nanoscience and nanoparticles based on their physical and electronic structures. This course was augmented with several lab activities described in section III. This course was taught by collaborators from Chemistry and Physics departments.

The second course, "Engineering Applications of Nanoscience & Nanotechnology" (cross-listed course numbers: EE490/590; ME490; Chem431; PHYS 492), was offered in spring semesters of 2016 and 2017. This course covered mechanical, electrical, and chemical properties of nanomaterials and the engineering applications of nanoscience and nanotechnology. It also covered ethical, social, and environmental impacts of nanomaterials. This course was taught by collaborators from Electrical, Chemical, and Mechanical Engineering departments.

To increase students' learning, two teaching methods were used: case studies and problem-based learning (PBL). These methods were well-suited for teaching prospective scientists and engineers because they focus on cooperative sharing of ideas as well as healthy discussion and resolution of problematic issues [7, 8]. PBL-structured case studies promote higher-order learning skills, such as application, analysis, synthesis, and evaluation. During case study-based learning modules, students were presented with a selected case to resolve the core issue by critically evaluating the information they had researched. They had opportunities to find the latest developments in a field and associate them with most recent social issues. This approach overcame the weakness of traditional lecture-based learning modules which may quickly become out of date for rapidly changing areas like NSNT without diligent attention from well-informed instructors. By its nature, PBL-structured case studies promote learning at the cutting edge of a discipline and thus are well-suited to the emerging NSNT field.

A central premise in using the case study technique is that the process of learning is just as important as the content [9]. In general, students work cooperatively during case studies to answer challenging questions or to evaluate complex ethical issues. For PBL-structured case studies, students are expected to investigate and learn necessary content in order to understand the context of a case. This requires students to become more involved in the learning process and help them develop traits that are important for mature scientists and engineers. The collaboration inherent in case studies presents a challenge for students who excel as individuals in lecture courses [10]. Cooperative problem solving is an essential approach that professional scientists and engineers utilize in their jobs, therefore it is important that undergraduate students are versed in this approach.

As part of this NSF project and to increase student learning, the first NSNT course was augmented with several lab activities. These activities aimed to introduce students to the optical properties of gold nanoparticles that can be changed dramatically as the size of the nanoparticles varies. Through these activities, students learned the difference in properties between nanomaterials and bulk materials. In addition, they learned how to use the Scanning Tunneling Microscope (STM) and the Atomic Force Microscopy (AFM) instruments acquired through the NSF funds. Examples of these lab activities are:

- Characterization of the nanoparticle using SEM
- Synthesis of gold nanoparticles
- Characterization of gold nanoparticles using SEM
- Characterization of gold nanoparticle optical properties using a spectrometer

## III    NSNT Nanotechnology Reflection Survey Descriptive Results

The project also explored students' motivations for investigating nanotechnology through interdisciplinary coursework. To collect this information, a NSNT Survey was administered to students who enrolled to two NSNT courses during the two years, 2016 and 2017.

A faculty member from the College of Education and Human Development visited each class on two occasions (at the beginning and end of the fifteen-week semester) to administer the Nanotechnology Reflection Survey. The Nanotechnology Reflection Survey [8] was created by Diefes-Dux and colleagues.

All students were given a statement of informed consent, emphasizing that participation in the data collection activity was voluntary. Students in the face-to-face, campus-based section of the course were given a hard copy of the consent form as well as the survey instrument. Students using the asynchronous, distance platform to take the course were given electronic versions of the documents. No compensation was offered to students for completing the survey. Response rates varied by semester (see table below). No distance students elected to complete the survey in any of the semesters.

| Semester | Beginning of Semester # of Responses | End of Semester # of Responses |
|---|---|---|
| Fall 2015 | 18 | 15 |
| Spring 2016 | 15 | 9 |
| Fall 2016 | 9 | 9 |
| Spring 2017 | 4 | 6 |

Students' self-reported levels of motivation for studying nanotechnology was measured across 17 items from the *Nanotechnology Reflection Survey*. All items used a five point scale. Response options included (1) strongly disagree, (2) disagree, (3) neutral, (4) agree, and (5) strongly agree. Results from students in the Nanoengineering and Nanoscience course (Fall 2015 and Fall 2016) are presented in Table 1 and Figure 1. Results from students in the Engineering Applications of Nanoscience & Nanotechnology course (Spring 2016 and Spring 2017) are presented in Table 2 and Figure 2.

Table 1: *Student Motivation for Studying Nanotechnology in an Introductory Nanoscience and Nanoengineering Course*

| I plan to: | Mean (beginning of course) | SD (beginning of course) | Mean (end of course) | SD (end of course) |
|---|---|---|---|---|
| Read a fiction story about nanotechnology. | 2.41 | 1.05 | 2.65 | 1.11 |
| Formally teach nanotechnology concepts (e.g. as a teaching assistant) | 2.44 | 0.97 | 2.87 | 0.92 |
| Investigate the implications of nanotechnology. | 4.15 | 0.60 | 4.00 | 0.90 |
| Informally/Casually teach someone something about nanotechnology. | 3.85 | 1.03 | 3.83 | 0.83 |
| Seek information about internships or Co-op experiences with companies engaged in nanotechnology. | 3.81 | 0.96 | 4.17 | 0.94 |
| Read a news story or popular magazine article about nanotechnology. | 3.96 | 1.02 | 4.09 | 0.79 |
| Give a presentation related to nanotechnology to an audience I perceived as having more experience with nanotechnology than I. | 3.07 | 1.07 | 3.22 | 0.90 |
| Explore nanotechnology as part of this course. | 4.63 | 0.49 | 4.26 | 0.86 |
| Read a research journal article about nanotechnology. | 4.04 | 0.94 | 4.04 | 0.98 |
| Enroll in a course about nanotechnology. | 4.48 | 0.58 | 4.36 | 0.73 |
| Attend a non-course related seminar about nanotechnology. | 3.46 | 0.95 | 3.48 | 1.20 |
| Visit an industry or business that specializes in nanotechnology. | 3.70 | 0.99 | 3.96 | 0.82 |
| Give a presentation related to nanotechnology to an audience I perceived as having less experience with nanotechnology than I. | 3.33 | 0.96 | 3.43 | 0.95 |
| Watch a program about nanotechnology. | 3.78 | 0.89 | 3.78 | 1.00 |
| Apply or interview for a nanotechnology related work or research experience. | 3.78 | 1.12 | 3.87 | 1.10 |
| Investigate fields of study in which I can learn more about nanotechnology. | 4.11 | 0.85 | 4.13 | 0.87 |
| Obtain a work experience or undergraduate research opportunity related to nanotechnology. | 3.81 | 1.08 | 3.87 | 0.81 |

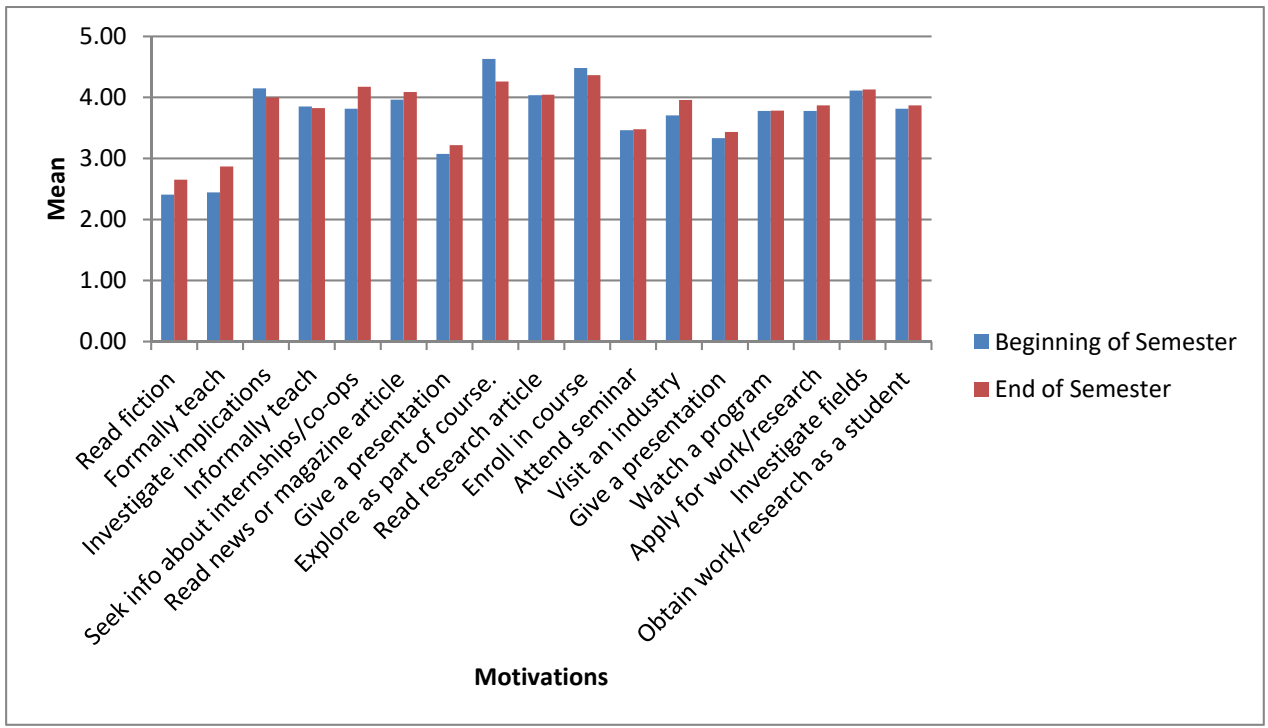

Figure 1: *Student Motivation for Studying Nanotechnology in an Introductory Nanoscience and Nanoengineering Course*

Table 2: *Student Motivation for Studying Nanotechnology in an Engineering Applications of Nanoscience and Nanotechnology Course*

| **I plan to:** | *Mean (beginning of course)* | *SD (beginning of course)* | *Mean (end of course)* | *SD (end of course)* |
|---|---|---|---|---|
| Read a fiction story about nanotechnology. | 2.68 | 1.00 | 2.73 | 0.80 |
| Formally teach nanotechnology concepts (e.g. as a teaching assistant) | 2.26 | 1.15 | 2.80 | 1.01 |
| Investigate the implications of nanotechnology. | 3.84 | 0.96 | 4.13 | 0.64 |
| Informally/Casually teach someone something about nanotechnology. | 3.89 | 0.88 | 4.07 | 0.80 |
| Seek information about internships or Co-op experiences with companies engaged in nanotechnology. | 3.53 | 1.12 | 3.73 | 0.80 |
| Read a news story or popular magazine article about nanotechnology. | 3.84 | 0.96 | 4.07 | 0.70 |
| Give a presentation related to nanotechnology to an audience I perceived as having more experience with nanotechnology than I. | 3.05 | 1.27 | 3.29 | 1.27 |
| Explore nanotechnology as part of this course. | 4.37 | 0.68 | 4.21 | 0.58 |
| Read a research journal article about nanotechnology. | 3.78 | 1.31 | 3.93 | 0.92 |
| Enroll in a course about nanotechnology. | 4.00 | 1.11 | 4.07 | 0.92 |
| Attend a non-course related seminar about nanotechnology. | 3.47 | 1.07 | 3.86 | 0.77 |
| Visit an industry or business that specializes in nanotechnology. | 3.79 | 1.13 | 4.07 | 0.83 |
| Give a presentation related to nanotechnology to an audience I perceived as having less experience with nanotechnology than I. | 3.26 | 1.24 | 3.43 | 1.09 |
| Watch a program about nanotechnology. | 3.58 | 1.12 | 3.57 | 1.02 |
| Apply or interview for a nanotechnology related work or research experience. | 3.47 | 0.96 | 3.93 | 0.83 |
| Investigate fields of study in which I can learn more about nanotechnology. | 3.68 | 1.06 | 3.92 | 0.64 |
| Obtain a work experience or undergraduate research opportunity related to nanotechnology. | 3.26 | 1.05 | 3.93 | 0.83 |

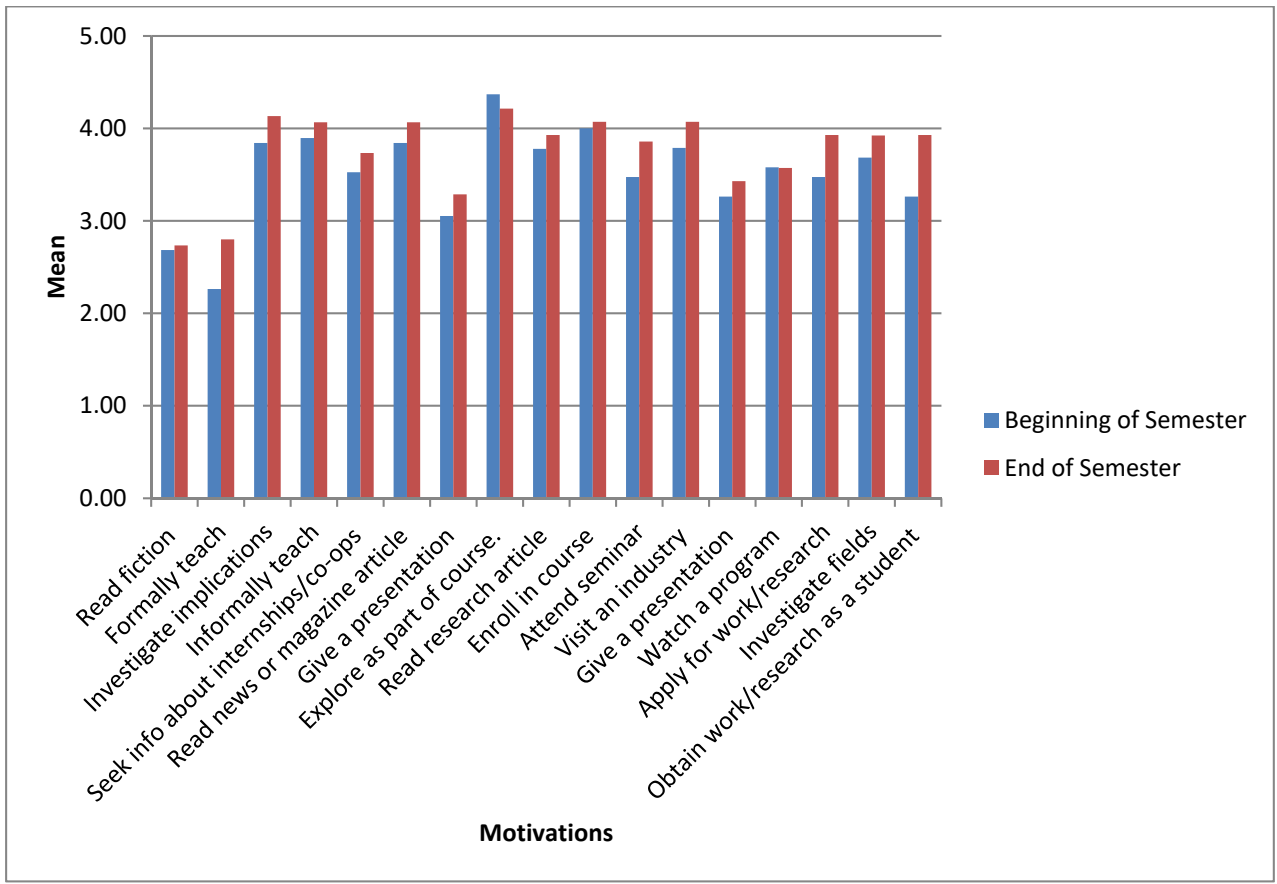

Figure 2: *Student Motivation for Studying Nanotechnology in an Engineering Applications of Nanoscience and Nanotechnology Course*

## IV    Discussion and Interpretation of Results

Students in the introductory Nanoscience and Nanotechnology courses demonstrated motivation to study nanotechnology by enrolling in the course and by exploring nanotechnology as part of the course. They also reported the opportunity to investigate the implications of nanotechnology as a motivator for studying nanotechnology in the first place. By the end of the courses, the students were motivated to seek information about internships or cooperative education work experiences with nanotechnology-related employers, and they also reported that they plan to read a news story or magazine article about nanotechnology in the future. There were no mean responses that indicated a significant decrease in plans to engage in activities related to nanotechnology from the beginning to the end of the course.

At the beginning of the courses on Engineering Applications of Nanoscience and Nanotechnology, it is not surprising that students reported the greatest motivators for studying nanotechnology being the class itself and the act of exploring nanotechnology through the course. By the end of the course, students were increasingly motivated to investigate the implications of nanotechnology and to teach others about what they had learned. They also expressed plans to read a news story or magazine article about nanotechnology, visit an industry of business that specializes in nanotechnology, and to obtain a work experience or undergraduate research opportunity related to nanotechnology.

The results concerning student motivation are encouraging to the development nanotechnology as a field of study and it seems logical that students would seek opportunities to apply what they learned after taking the Nanoscience and Nanotechnology course as well as the Engineering Applications of Nanoscience and Nanotechnology course. In general, students expressed intent to follow formal classroom learning about nanotechnology with out-of-class experiences that were experiential in nature. The act of learning by doing serves to reinforce concepts to which they were exposed in the classroom and the courses, particularly the second course related to engineering applications of nanoscience and nanotechnology, provided students with the knowledge base to do so.

## V    Conclusions

In this paper, a NSF-funded project that aims to establish a nanoscience and nanotechnology program at the University of North Dakota was described. The project's aims are to foster new interest in nanoscience and nanotechnology among UND STEM students, provide them with the knowledge and skills necessary for the NSNT field, and prepare them to become the next generation of graduates to compete in the global market. In addition to courses, the collaborators developed and integrated hands-on activities in these courses for STEM students. Furthermore, Native American college and high school students participating in the NATURE program, were exposed to nanoscience and nanotechnology through the Sunday Academy program.

**Acknowledgements**

The authors acknowledge the support of the National Science Foundation (NSF), grant # 1443861.